\documentclass[prl,english,showpacs,nofootinbib, notitlepage, twocolumn,superscriptaddress]{revtex4-2}
\usepackage{amssymb}
\usepackage{color}
\usepackage{graphicx}
\usepackage{bbold}
\usepackage{bm}
\usepackage{amsmath}
\usepackage{amsfonts}
\usepackage{amssymb}
\usepackage{braket}
\usepackage{units}
\usepackage{csquotes}
\usepackage{upgreek}
\usepackage{xcolor}
\usepackage{fancyhdr}
\usepackage{float}
\usepackage{lipsum}

\begin{document}
	

\title{Detecting the phase transition in a strongly-interacting Fermi gas by unsupervised machine learning}

\author{D.~Eberz}
\affiliation{Physikalisches Institut, University of Bonn, Wegelerstra{\ss}e 8, 53115 Bonn, Germany}
\author{M. Link}
\affiliation{Physikalisches Institut, University of Bonn, Wegelerstra{\ss}e 8, 53115 Bonn, Germany}
\author{A. Kell}
\affiliation{Physikalisches Institut, University of Bonn, Wegelerstra{\ss}e 8, 53115 Bonn, Germany}
\author{M. Breyer}
\affiliation{Physikalisches Institut, University of Bonn, Wegelerstra{\ss}e 8, 53115 Bonn, Germany}
\author{K. Gao}
\affiliation{Physikalisches Institut, University of Bonn, Wegelerstra{\ss}e 8, 53115 Bonn, Germany}
\affiliation{Department of Physics, Renmin University of China, Beijing 100872, China}
\author{M. Köhl}
\affiliation{Physikalisches Institut, University of Bonn, Wegelerstra{\ss}e 8, 53115 Bonn, Germany}

\begin{abstract}

{ We study the critical temperature of the superfluid phase transition of strongly-interacting fermions in the crossover regime between a Bardeen-Cooper-Schrieffer (BCS) superconductor and a Bose-Einstein condensate (BEC) of dimers. To this end, we employ the technique of unsupervised machine learning using an autoencoder neural network which we directly apply to time-of-flight images of the fermions. We extract the critical temperature of the phase transition from trend changes in the data distribution revealed in the latent space of the autoencoder bottleneck. }
\end{abstract}
\maketitle

An ensemble of attractively-interacting fermions exhibits a phase transition to a superfluid state below a critical temperature $T_\mathrm{C}$. The exact temperature at which the phase transition occurs depends on the microscopic details, such as inter-particle interactions and correlations. For weak attractive interactions, in the Bardeen-Cooper-Schrieffer (BCS) regime, the phase transition is governed by the opening of a gap due to Cooper instability near the Fermi level. The critical temperature in this regime decays exponentially with decreasing interaction strength. If the system supports a dimer bound state between two fermions, the ensemble can form a molecular Bose-Einstein condensate (BEC). The critical temperature of this state converges towards the value of a weakly repulsive BEC for decreasing interaction strength. These two regimes are known as the limits of the BEC-BCS crossover, connected by the unitarity regime around the point of diverging scattering length. In the regime of strong interaction strength around unitarity the determination of the critical temperature is a field of ongoing research \cite{sa1993, haussmann2007, bulgac2008, burovski2008, inada2008, floerchinger2008, floerchinger2010, pisani2018, pini2019}.

Detecting the phase transition over a wide range of interactions has been difficult. Only on the BEC side of the crossover, the conventional technique of detecting the bimodal momentum distribution of the dimers directly reveals the condensate \cite{greiner2003, zwierlein2003, jochim2003}. In contrast, at unitarity a measurement of the equation of state and thermodynamic quantities unveiled the critical temperature \cite{ku2012}. On the BCS side of the crossover, the Cooper pairs break upon release from the trap and, therefore, the so-called rapid-ramp technique has been developed to convert these pairs to tightly-bound dimers \cite{regal2004, zwierlein2004, altman2005, tikhonenkov2006}. Whether or not the rapid-ramp technique closely reflects the situation of the trapped gas depends crucially on the adiabaticity of the ramp and an accurate verification of this is very difficult. A direct detection of the superfluid signature in the momentum distribution of the fermions, on the other hand, is obscured by finite temperature, collisions during ballistic expansion and the shape of the trapping potential. Recently, we have demonstrated that using supervised learning of deep neural networks, the condensate fraction and hence the critical temperature over a wide range of interactions can be detected directly from the momentum distribution of the fermions \cite{link2022}. However, this method still relies on the rapid-ramp technique for labelling the training data.

\begin{figure}
	\includegraphics[width=0.5 \textwidth]{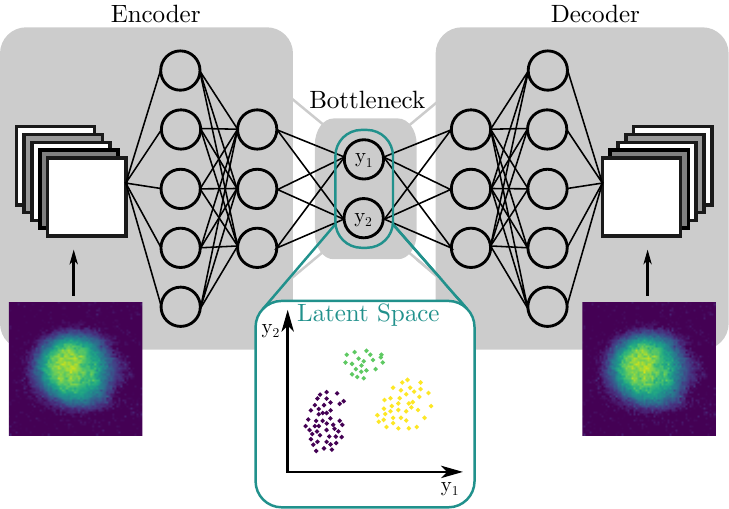}
	\caption{Architecture and training of an autoencoder network. By keeping the number of neurons in the bottleneck low and training the network to reproduce its input, a low-dimensional representation of the input data can be generated. The output of the bottleneck can then be accessed to search for features in the data structure. \label{fig1}}
\end{figure}

In this work, we measure the critical temperature of the superfluid phase transition by employing unsupervised machine learning directly on time-of-flight images. Unsupervised machine learning is a technique which does not require labelling the data during training of the network and hence is unbiased. This is a significant advantage because the generation of labels is a potential source of error. 
To this end, we employ a deep neural network as an autoencoder as illustrated in Figure \ref{fig1}. The autoencoder comprises an encoder and decoder network as well as a bottleneck layer in the middle, with the input of the encoder being of the same dimension and shape as the output of the decoder. The bottleneck layer connects the input and output layers and, preferably, has much lower dimensionality than both input and output. Further, the encoder neural network is trained to compress the features of the input data (e.g.\ a picture) to the few neurons of the bottleneck, while the decoder neural network uses the bottleneck as an input to replicate the original input signal as accurately as possible. In this approach, a low-dimensional bottleneck will enforce an efficient representation of the data into a few meaningful parameters. 
If the autoencoder is successful, the latent space of the neuron outputs in the bottleneck contains relevant information, which can be interpreted to classify the input data. This is especially noteworthy because, by construction, the autoencoder does not take any additional information other than the raw data into account. 
So far, unsupervised machine learning has been used on theoretical data with learning by ``confusion'' \cite{nieuwenburg2017}, principal component analysis \cite{wang2016, nieuwenburg2017, hu2017}, autoencoders \cite{hu2017, alexandrou2020}, and for the classification of topological phases in experimental data \cite{kaming2021}.

Experimentally, we prepare a quantum gas of $\sim 3\cdot10^{5}$ atoms per spin state in the two lowest hyperfine states $\ket{1}$ and $\ket{2}$ of ${}^{6}\textrm{Li}$ in a crossed-beam optical dipole trap \cite{behrle2018}. We adjust the interaction strength of the sample by utilising a magnetically controlled Feshbach resonance and the temperature by controlled heating from a time-dependent variation of the trapping potential \cite{link2022}. 

This approach enables us to tune the interaction strength and temperature independently of each other. In addition, we measure the central density and temperature of the cloud by performing an inverse Abel transform to the column density of in-situ images. After reconstructing the 3D density distribution in the trap, we use the central density to calculate the homogeneous Fermi energy and fit a virial expansion of the equation of state to the edge of the cloud to determine the temperature. The thermalised cloud is detected by absorption imaging after $5\,\text{ms}$ of time-of-flight, which serves as the raw data for the autoencoder analysis. In the ana\-lysis, the calibration is used to assign time-of-flight images of variable heating time with a corresponding Fermi energy and temperature. Further details regarding the detection and calibration of the data are given in our previous work, which shares the same data set \cite{link2022}.

In total, we accumulate $5031$ time-of-flight pictures at different interaction strengths and temperatures (see Figure \ref{fig1}), which we split into $90\%$ for training and $10\%$ for validation after random shuffling. The network is written within the TensorFlow framework \cite{tensorflow} and trained with stochastic gradient descent via the Adam optimiser \cite{adam}. The architecture of the autoencoder resembles two mirrored convolutional neural networks connected by a two-dimensional bottleneck layer. Both networks have independent weights and are trained simultaneously in form of the autoencoder to reconstruct the time-of-flight images. Later, the outputs of the bottleneck are extracted to classify the data. The details of the network architecture are given in the Appendix. 

\begin{figure}
\includegraphics[width=0.5\textwidth]{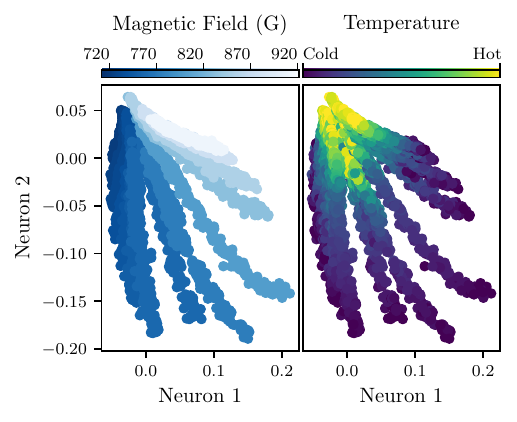}
\caption{Latent space representation of time-of-flight data across the crossover, in which every data point represents one compressed image. \textbf{(Left)} The colour of the data indicates the external magnetic field strength. In the latent space, the data is organized along curves of sorted interaction values. \textbf{(Right)} Same plot as (Left), but the colour shows the relative temperature of data points. Since the temperatures differ for each field the scale is normalized. The network not only learns to sort data points by interaction but also arranges them by temperature along each curve, with the tendency of hot clouds to converge towards a common region in the latent space. \label{fig2}}
\end{figure}

After training, we use the encoder and bottleneck parts of the network to achieve the two-dimensional latent space representation shown in Figure \ref{fig2}. The left panel of the figure shows the latent space spanned by the two bottleneck neurons with a colour code corresponding to the magnetic field strength which determines the interaction strength. We observe that all data belonging to the same magnetic field lie along a curve in the latent space. The right panel shows the same representation of the latent space, however, with the colour map representing temperature. This depiction illustrates that the ordering along the lines of different interaction strengths is according to the temperature of the sample. It should be noted that the autoencoder does not receive any information regarding the interaction or temperature, but learns these quantities from the time-of-flight pictures alone. Additionally, the autoencoder groups thermal (hot) gases in a small concentrated area in the upper-left corner, while the (cold) superfluid gas covers a larger area at the bottom of the latent space. Hence, the neural network detects a wider range of variation in the signatures of superfluids for different interaction strengths, whereas differently interacting thermal gases are more similar to each other. For the work discussed in this paper, the reduction to two neurons at the bottleneck has proven sufficient. We found that the final loss during training, which measures the mean squared error of the network outputs with regard to the inputs, shows no improvement for more than two neurons. Moreover, for trained networks with three bottleneck neurons the output in the latent space is arranged on a common plane reducing the latent space to a two-dimensional distribution.  

\begin{figure}
	\includegraphics[width=0.5 \textwidth]{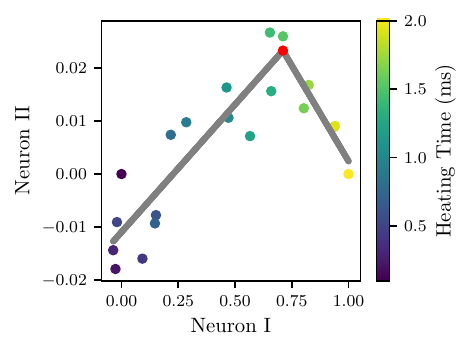}
	\caption{Extraction of the critical temperature from the latent space representation. As an example the extraction for data with $1/(k_{\textrm{F}}a) \sim -0.26$ at the trend changing position is shown. Each point represents the averaged position of $13$ recorded images. The grey lines display the fitted piecewise linear segments and the red point shows the determined position of the trend change. \label{fig3}}
\end{figure}

Tracing the data points from hot to cold for a given interaction strength shows that the data is not well described by a single straight line. Instead, we use piecewise linear segments to model the data and to detect a change of trend at specific temperatures. First, we average the latent space position of the $\sim 13$ recorded images for each temperature and interaction strength setting. Next, we trim the considered temperatures for each interaction strength to a range around the determined critical temperature from our supervised analysis \cite{link2022}. This approach prevents the accidental detection of trend changes at tempera\-tures far above or below the critical temperature. Around unitarity where the reduced temperature $T/T_\mathrm{F}$ does not change substantially at the expected critical temperature, the trimming range is chosen such that we omit temperatures at which the condensate fraction from our supervised analysis is above $\sim 5\,\%$. At fields where the reduced temperature varies strongly with the determined critical heating time and the maximally measured condensate fraction is below this limit, the chosen range comprises temperatures in a window of roughly $\pm 25\,\% \cdot T/T_\mathrm{F}$ of the previously determined critical temperatures. This range covers approximately $80 \textrm{--} 100\,\%$ of the span of measured temperatures at each field. Subsequently, the latent space is shifted and linearly transformed, while preserving angles, before we apply a piecewise linear fit in order to determine the location of the trend change as shown in the transformed latent space (Neuron I and II) in Figure \ref{fig3}. The fit is performed by defining a piecewise linear function with four degrees of freedoms namely two slopes and the coordinates of the trend change which is fitted to the data via the least squares method. We average the heating time of the three data points closest to the trend changing position, and use our calibration of the heating measurements to infer the corresponding temperature. Since the trained network model is subject to statistical variation caused by the finite size of the shuffled data set, stochastic gradient descent during training and randomly initialised weights of the network, we train multiple equivalent models. Next, we average the determined heating times at the trend changing positions weighted by the reciprocal variance of the previously calculated mean of the three closest data points. In total we train $173$ models and omit only one model with non-converged loss after training. In addition, we omit every fit in which the result is at the edge of the trimmed latent space, which sums up to $33$ discarded fits from a total of $2408$ for all of the $172$ remaining models with $14$ different interactions each. These fits mainly belong to interactions around $1/(k_\mathrm{F} a) \sim 0.5$, where the change of trend is least pronounced.

\begin{figure}[h!]
	\includegraphics[width=0.5 \textwidth]{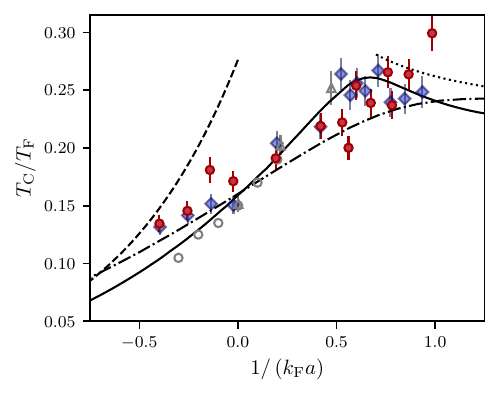}
	\caption{Extracted critical temperatures across the BEC-BCS crossover shown as red dots. Error bars are calculated from the SE of averaging multiple trained models with differently shuffled training data sets and an estimation of the systematic error caused by non-harmonicities of the trap \cite{link2022}. Blue diamonds represent the results of our supervised analysis \cite{link2022}. Dashed line: BCS theory with GMB corrections; solid line: extended GMB	theory \cite{pisani2018}; dash-dotted line: theory from \cite{haussmann2007}; dotted line: interacting BEC; open triangles: quantum Monte-Carlo data \cite{burovski2008}, open circles: quantum Monte-Carlo data \cite{bulgac2008}. \label{fig4}}
\end{figure}

In Figure \ref{fig4} we compare the extracted critical tempera\-tures to the phase boundary determined by our supervised neural network \cite{link2022} and several theories \cite{haussmann2007, bulgac2008, burovski2008, pisani2018}, finding overall very good agreement. Our analysis reveals the critical temperature in the region of $-0.40 < 1/(k_\mathrm{F} a) < 0.98$ and shows striking resemblance with the extended Gor'kov-Melik-Barkhudarov (GMB) theory \cite{pisani2018}. On the BCS side this approach agrees well with the higher temperatures presented by reference \cite{haussmann2007}, which exceed the suggested temperatures by Quantum-Monte Carlo (QMC) calculations \cite{bulgac2008}. From unitarity to the BEC side of the crossover the temperatures match with the QMC calculations in \cite{burovski2008}. Moreover, the determined superfluid transition temperatures agree well with the values from our previous work using supervised training of deep neural networks \cite{link2022}. We therefore conclude that the change of trend in the latent space corresponds to the critical temperature, providing a new and unbiased perspective on the phase boundary in the crossover regime. Thus, this work presents another confirmation of our previous work, which in contrast does not rely on the rapid-ramp technique while retaining good accuracy.

In conclusion, we extracted the phase boundary of the superfluid phase transition in the BEC-BCS crossover directly from the momentum distribution of the fermions by application of an unsupervised autoencoder network. The network is able to identify physical quantities like temperature and interaction strength independently of any external information, and reveals the phase transition by a feature in the low-dimensional latent space of the data set. Our work shows that the direct detection of Fermi condensates from the momentum distribution of fermions is possible even without projection onto dimers. In this regard, this method presents a new approach to unveil features in experimental time-of-flight data where model-based analysis is impractical.

This work has been supported by the DFG (SFB/TR 185 project C6) and Cluster of Excellence Matter and Light for Quantum Computing (ML4Q) EXC 2004/1 – 390534769. 

\section*{APPENDIX}

\subsection*{Autoencoder network}

The neural network architecture used to create the low-dimensional representation of the phase diagram is given in Table \ref{tab1:autoencoder_architecture}. For training we provide the network with $5031$ centred absorption images of $192 \times 192\,\mathrm{px}$ at different temperatures and interactions across the crossover with $\sim 13$ repetitions each. To generate the latent space the full network is trained in a supervised way using the whole data set as both input and labels at the same time. Later, only the encoder half of the network is employed to extract the latent space representation of the data set in the bottleneck layer. This approach neither requires the network to share a symmetric encoder and decoder, nor to have shared weights of mirrored layers. Every dense and convolutional layer uses the ReLU activation, except for the dense layer in the bottleneck utilising a linear activation. The total number of parameters is $17400963$ and the None in the output shape describes the batch dimension. We used a batch size of 20 and trained for 15 epochs with Adam optimiser \cite{adam} and a learning rate of $4\times10^{-4}$ with mean squared error as loss function. From the $5031$ data points we use $90\,\%$ for training and $10\,\%$ for validation, which are randomly shuffled before being used for training. The network was realised with the TensorFlow library \cite{tensorflow}.

\begin{table}[!htb]
\centering
\caption{Neural network architecture used for the autoencoder.}
\label{tab1:autoencoder_architecture}
\begin{tabular}{|c|c|c|}
\hline
Layer (type) & Output Shape & Parameters\\ \hline
Input & (192, 192, 1) & 0 \\ \hline
2D Convolutional & (192, 192, 32) & 320 \\ \hline
Max Pooling & (48, 48, 32) & 0 \\ \hline
2D Convolutional & (48, 48, 64) & 8256 \\ \hline
Max Pooling & (16, 16, 64) & 0 \\ \hline
Flatten & (16384) & 0 \\ \hline
Dense & (512) & 8389120 \\ \hline
Dense & (512) & 262656 \\ \hline
Dense (Bottleneck) & (2) & 1026 \\ \hline
Dense & (512) & 1536 \\ \hline
Dense & (512) & 262656 \\ \hline
Dense & (16384) & 8404992 \\ \hline
Reshape & (16, 16, 64) & 0 \\ \hline
2D Transpose Convolutional & (48, 48, 64) & 36928 \\ \hline
Batch Normalization & ( 48, 48, 64) & 256 \\ \hline
2D Transpose Convolutional & (192, 192, 32) & 32800 \\ \hline
Batch Normalization & (192, 192, 32) & 128 \\ \hline
2D Convolutional & (192, 192, 1) & 289 \\ \hline
\end{tabular}
\end{table}

\end{document}